\documentclass[10pt]{article}          %%
\usepackage{mathrsfs,amsfonts}
minus.4em \belowdisplayshortskip=2mm plus.2em minus.4em

\makeatletter 
\@addtoreset{equation}{section}  \makeatother

\title{\centerline{\normalsize } %\hfill hep-th/0604188}
\bf Asymptotic quasinormal  modes of a noncommutative geometry inspired
Schwarzschild black hole}

\author{ {\bf Pulak Ranjan Giri\thanks{e-mail :pulakranjan.giri@saha.ac.in}}\\
\normalsize Saha Institute of Nuclear Physics, 1/AF Bidhan-Nagar, Calcutta
700064, India}

\date{\today}

\begin{document}\maketitle
%%%%%%%%%%%%%%%%%%%%%%%%%
\begin{abstract} \noindent\small
We study the  asymptotic quasi-normal modes  for the scalar perturbation of the
non-commutative geometry inspired Schwarzschild black hole in (3+1)
dimensions. We have considered  $M\geq M_0$, which effectively correspond to a
single horizon Schwarzschild black hole with correction due to
non-commutativity. We have shown that for this situation the real part of the
asymptotic quasi-normal frequency is proportional to $\ln (3)$. The effect of
non-commutativity of spacetime on quasi-normal frequency arises through the
constant of proportionality, which is Hawking temperature $T_H(\theta)$. We
also consider the two horizon case and show that in this case also the real
part of the  asymptotic quasi-normal frequency is proportional to $\ln(3)$.
\end{abstract}
%%%%%%%%%%%%%%%%%%%%%%%%%
\hspace{1cm}
%Pacs:~{04.30.-w, 04.60.-m, 04.70.-s, 04.70.Dy}

%%%%%%%%%%%%%%%%%%%%%%%%%
\section{\small{\bf {Introduction}}} \label{in}
%%%%%%%%%%%%%%%%%%%%%%%%%
Black hole spacetime, when perturbed by external fields, the perturbations are
radiated away and the black hole spacetime returns to its equilibrium. The
radiated waves, called quasi-normal modes \cite{press1,press2,press3} 
are characterized by
complex frequency $\omega$. They depend only on the parameters of the black
hole spacetime and are independent of the details concerning the initial
perturbation. On the other hand asymptotic quasi-normal modes (quasi-normal
modes which damp infinitely fast) do not radiate at all and therefore they can
be interpreted as the fundamental oscillation for the black hole spacetime.

Asymptotic quasi-normal modes are supposed to have a role to play in the quest
for a theory of quantum gravity \cite {hod,dreyer}. It is further suggested
\cite{dreyer} that asymptotic quasi-normal frequency may help to fix certain
parameters in loop quantum gravity. The key factor which makes people believe
that asymptotic quasi-normal modes have important role in quantum gravity is
the real part of the asymptotic quasi-normal frequency, which was supposed to
have the form $Re(\omega)\sim\ln k$ ($k$ is the natural number). But this fact
is seen to be  true for $d$ dimensional ($d>3$) Schwarzschild black holes
\cite{motl1,motl2}, not for all black hole spacetimes. For example, for four
dimensional Reissner-Nordstr\"{o}m black hole, the asymptotic quasi-normal
frequencies  have complicated form \cite{motl2}, which is not of the desired
from $Re(\omega)\sim\ln k$. For four dimensional Schwarzschild de Sitter
$(dS)$ and Schwarzschild Anti-de Sitter $(AdS)$ black hole, again the real
part of the asymptotic quasi-normal frequency \cite{cardoso1} does not have the
form $Re(\omega)\sim\ln k$. So, the important question of universality of the
real part of the asymptotic quasi-normal frequency of black hole as suggested
in \cite{hod,dreyer} is evidently not valid
\cite{cardoso2,cardoso3,cardoso4}. 
Schwarzschild spacetime is in active investigation \cite{ko1,ko2,ko3} 
for many years.

But all the studies have been performed  for the  black holes, where spacetime
are commutative. Non-commutative spacetime, on the other hand gets special
interest due to the prediction of string theory 
\cite{witten1,witten2,witten3,witten4,witten5}, which along
with the brane-world scenario \cite{anto}, lead to the fact that spacetime
could be non-commutative. One more important thing about studying
non-commutative geometry is probably the various kind of divergence, which
appear in General Relativity. It is believed that spacetime non-commutativity
may cure this divergence in General Relativity. The analysis of  (1+1) and
(3+1) dimensional linearized General Relativity \cite{gru1,gru2} reinforce this
belief. Schwarzschild black hole has been studied taking non-commutative space
into account \cite{nasseri1,nasseri2,nasseri3} and it has been shown that 
to the first order of
the non-commutative parameter $\theta$ there is no effect in the spacetime
metric of the black hole, but in second order there is a effect of
non-commutative space in the metric if the linear momentum in the plane
perpendicular to the non-commutative parameter is not zero. On the other hand
Schwarzschild black hole in non-commutative spacetime has been studied in
\cite{nicol,nicoli}. It has been shown that the modified metric due to
non-commutativity of spacetime does not allow the black hole to decay beyond a
critical mass $M_0= 1.9 \sqrt{\theta}$ through Hawking radiation  and Hawking
temperature does not diverge at all, rather it reaches a maximum value before
cooling down to absolute zero.

Having said so much about the non-commutative black hole, it naturally comes to
mind that, how does  non-commutativity affect the quasi-normal modes, especially
to the real part of the asymptotic quasi-normal frequency? It's an open
question, and so far has not been answered. In our present work, we are going
to address  this issue in detail. We will calculate the asymptotic quasi-normal
modes for the non-commutative  geometry inspired Schwarzschild black hole
\cite{nicol}, where non-commutativity can be taken as the correction to the
Schwarzschild black hole metric and goes to zero when the strength of
noncommmutativity goes to zero. We will show that although the geometry is
modified in presence of non-commutativity, the real part of the asymptotic
quasi-normal frequency remains proportional to $\ln (3)$.

We have organized the paper as follows: In Sec. (2), we  discuss the
non-commutative geometry inspired Schwarzschild black hole \cite{nicol}. In
Sec. (3),  we  find out analytically the asymptotic quasi-normal modes  and
calculate the asymptotic quasi-normal frequency of the single horizon black
hole spacetime. In Sec. (4), we find out analytically the asymptotic
quasi-normal modes  and calculate the asymptotic quasi-normal frequency of the
two horizon black hole spacetime. In Sec. (5), we conclude with some
discussion.

%%%%%%%%%%%%%%%%%%%%%%%%%
\section{\small{\bf{Non-commutative geometry inspired Schwarzschild black hole}}}\label{so}
%%%%%%%%%%%%%%%%%%%%%%%%%%
The Schwarzschild black hole is usually modified  when non-commutative
spacetime is taken into account  and the modified metric is given by (in units
of $G_N=1, c=1$) \cite{nicol}
\begin{equation}
 ds^2 = - f(r) dt^2 + f(r)^{-1} dr^2 + r^2\,\left(\, d\vartheta^2
+\sin^2\vartheta\, d\phi^2\,\right), \label{metric}
\end{equation}
with
\begin{equation}
f(r)=  1- \frac{4M}{r\sqrt{\pi}}\, \gamma(3/2 \ , r^2/4\theta\,)
\label{coef}
\end{equation}
where $r,\vartheta,\phi$ are the spherical coordinates, $\theta$ is the
strength of  non-commutativity of spacetime and $M$ is the mass of the black
hole. The upper incomplete Gamma function $\gamma\left(3/2 \,, r^2/4\theta\,
\right)$ is given by:

\begin{equation}
\gamma\left(3/2\ , r^2/4\theta\, \right)\equiv \int_0^{r^2/4\theta} dt\,
t^{1/2} e^{-t}\,.
\end{equation}
The horizons can be found from the equation,
\begin{equation}
f(r)=1- \frac{4M}{r\sqrt{\pi}}\, \gamma(3/2 \ , r^2/4\theta\,)\,=0\,.
\end{equation}
But it is impossible to solve this equation analytically in exact
form. However,  it is possible to solve it numerically \cite{nicol} and get
the radius of the horizons. Numerical solution shows that it has different
horizons for different values of the mass $M$ of the black hole:
\begin{enumerate}
\item For $M> M_0= 1.9\times (\sqrt\theta) /G$, there is two distinct horizons;
\item For $M = M_0= 1.9\times (\sqrt\theta) /G$, there is one degenerate
horizon at $r_0=3.0\times \sqrt\theta $, which corresponds to extremal black
hole;
\item For $M< M_0= 1.9\times (\sqrt\theta) /G$, there is no horizon.
\end{enumerate}
Our present interest is for the case of two horizons, because by increasing
the mass $M$ of the black hole, i.e. for $M\gg M_0$, the inner horizon can be
shrunk to zero and the outer horizon then approaches to Schwarzschild value
$r_h= 2M$. So effectively we can consider it as  a single horizon
Schwarzschild black hole, where the event horizon is modified due to
non-commutativity. The event horizon for this case can be written in terms of
upper incomplete gamma function
\begin{equation}
 r_h \approx 2M\,\left[\, 1 -\frac{2}{\sqrt{\pi}}\, \gamma\left(\, 3/2\ ,
M^2/\theta\, \right)  \,\right]\,, \label{horizon}
\end{equation}
and the Hawking temperature for the black hole is given by

\begin{eqnarray}
T_h(\theta)&&\equiv \,\frac{1}{4\pi}\,\left[ \frac{d f(r)}{dr}\right]_{r=r_h}
\nonumber\\ &&= \frac{1}{4\pi\,r_h}\left[\, 1 -\frac{r^3_h}{4\,\theta^{3/2}}\,
\frac{e^{-r^2_h/4\theta}}{\gamma\left(\, 3/2\,, r^2_h/4\theta \right)}
\,\right]\,. \label{tem}
\end{eqnarray}
It is easy to see that for $\theta \to 0$ Eq. (\ref{horizon}) and Eq.
(\ref{tem}) reduces to the pure Schwarzschild value.
%%%%%%%%%%%%%%%%%%%%%%%%%%%%%%%%%%%%%%%%%%%%%%%%%%%%%
\section{\small{\bf{Asymptotic quasi-normal mode analysis for the single
      horizon  black hole}}}\label{an}
%%%%%%%%%%%%%%%%%%%%%%%%%%%%%%%%%%%%%%%%%%%%%%%%%%%%%%%
In this section we obtain asymptotic  quasi-normal frequency corresponding to
the scalar perturbations for spherically symmetric (3+1) dimensional
non-commutative geometry inspired Schwarzschild black-hole \cite{nicol}
discussed in previous section. Before proceeding further one important point we
would like to mention is that, in \cite{nicol}, the effect of non-commutativity
has been introduced in the right hand side of the Einstein equation  as
modified matter distribution keeping the Einstein tensor in the left hand side
intact. In this same spirit we also consider the Klein Gordon equation on
curved spacetime, where the metric is taken from Ref. \cite{nicol} and we have
kept everything except the metric like commutative spacetime. Once we assume
this, then the evolution equation for the scalar field perturbation follows
directly from the massless, minimally coupled scalar field propagating in the
line element Eq. (\ref{metric}). The scalar perturbation is governed by,
\begin{equation}
\Box~\Phi \equiv \frac{1}{\sqrt{-g}}~ \partial_{\mu} \left(
\sqrt{-g}g^{\mu\nu}\partial_{\nu} \Phi \right) =  0 \,.
\end{equation}
Now with the trial solution of the type:
\begin{equation}
\Phi(x^{\mu}) = \frac{1}{r} \, R(r) \, \exp(i \omega t) \, Y_{l
m}(\vartheta,\Phi)\,,
\end{equation}
we can separate the radial equation, which in tortoise coordinate $x=
\int\frac{dr}{f(r)}$ is of the form,
\begin{equation}
\frac{d^2 R(r)}{dx^2} + \left[ \omega^2 - V(r) \right] R(r) = 0 \, ,
\label{dr}
\end{equation}
where $V(r)$ is the Regge-Wheeler potential \cite{regge} and is given by,
%
%%%%%%%%%%%%%%%%%%%%%%%%%%%%%%%%%%%%%%%%%%%%%
\begin{equation}
V(r) =  f(r)\left[\frac{l(l+1)}{r^2} + \frac{1}{r}\frac{d}{dr} f(r) \right]\,
. \label{rege}
\end{equation}
In our analysis we would use the monodromy method of ref.  \cite{motl2}. We
need to extend (\ref{dr}) throughout the entire complex  $r$ plain in order to
use monodromy technique \cite{motl2}.

The boundary condition satisfied by the scalar field  $\Phi (x)$ is given by
\begin{equation}
\Phi (x) \sim e^{i\omega x}\;\, {\mathrm{as}}\;\, x \to - \infty,
\label{bc1}\\ \Phi (x) \sim  e^{-i\omega x}\;\, {\mathrm{as}}\;\, x \to +
\infty.\label{bcondition}
\end{equation}

Near the origin  $r=0$, the tortoise coordinate looks like
\begin{equation}
 x \sim -\frac{\Gamma(3/2)r^{2}}{4
 M\gamma(3/2,r_0^2/4\theta)}. \label{tortoise}
\end{equation}
Here we have assumed  that $r^2/4\theta$ is sufficiently large near origin
such that $\frac{\gamma(3/2,r^2/4\theta)}{\Gamma(3/2)}$ can safely  be taken
outside the integral, when performing the integration for evaluation of
$x$. We can do it by suitably adjusting the non-commutative parameter
$\theta$. $r_0$ is any point near origin, which we have put to keep it outside
the integral. The potential $V(r)$ near the origin takes the form,
\begin{equation}
V(r(x))=\frac{j^2-1}{4x^2}, \label{pot}
\end{equation}
where $j= 0$. For highly damped asymptotic quasi-normal modes, we take the
frequency $\omega$ to be approximately purely imaginary.  Thus, for the Stokes
line defined by ${\rm Im}~ (\omega x) = 0$, $x$ is approximately purely
imaginary. This together with Eq.  (\ref{tortoise}) implies that near $r=0$,
the behavior of $r$ is of the form
\begin{equation}
r= \rho~e^{i\pi/4}e^{in\pi/2}
\end{equation}
with $\rho > 0$ and $n= 0,1,2$ and $3$.  The signs of $\omega x$ on these
 lines are given by $(-1)^{n}$ and near the origin, the Stokes lines are
 equispaced by an angle $\frac{\pi}{2}$. Also note that near infinity, $x \sim
 r$ and ${\rm Re}~(x) = 0$ and ${\rm Re}~(r) = 0 $ are approximately
 parallel. Two of the stokes lines are parallel to the two imaginary $r$ axis
 and other two Stokes lines starting from the origin would form a closed loop
 in the complex $r$ plane \cite{motl2}.

Now to calculate quasi-normal modes, we will consider the contour of figure (2)
drawn in complex $r$ plane in Ref. \cite{motl2}. The reason why we are
considering the figure (2) of  Ref. \cite{motl2} for our calculation needs to
be clarified here. In our analysis of quasi-normal modes as we have said
earlier that we are considering the case  $M\geq M_0$. Although for $M>M_0$,
the black hole \cite{nicol} has two horizons, we can consider it as a single
horizon black hole when mass of the black hole $M$ becomes much much larger
than the critical mass $M_0$. So in this situation the non-commutative geometry
inspired Schwarzschild black hole spacetime geometry is similar to the
Schwarzschild geometry, except the modified horizon. The solution of the wave
Eqn. (\ref{dr}) near origin is given by
\begin{equation}
\Phi(x) =  A_+ \sqrt{2\pi\omega x}J_{\frac{j}{2}}(\omega x)+ B_+
\sqrt{2\pi\omega x}J_{-\frac{j}{2}}(\omega x),\label{phi}
\end{equation}
where $J_{\nu = {\pm\frac{j}{2}}} $ are the Bessel functions of first kind and
$A_+,~B_+$ are constants. Since we are considering the situation where ${\rm
Im}~(\omega) \rightarrow \infty$, we can use the asymptotic expansion of the
Bessel function to write the solution  Eq. (\ref{phi}) as
\begin{equation}
\Phi (x) = \left( A_+ e^{-i\alpha_+} + B_+ e^{-i\alpha_-}\right) e^{i \omega
x} + \left( A_+ e^{i\alpha_+} + B_+ e^{i\alpha_-}\right) e^{-i \omega x},
\label{solution}
\end{equation}
where $\alpha_\pm = \frac{\pi}4 (1 \pm j)$. Now consider a region near $A$ (
$A$ is a point indicated in figure (2) of Ref.  \cite{motl2}), where we have
$\omega x \rightarrow \infty$ and $x= \rightarrow +\infty$. So imposing the
boundary condition Eq.  (\ref{bcondition}), we get from Eq.(\ref{solution}):
\begin{equation}
 A_+ e^{-i\alpha_+} + B_+ e^{-i\alpha_-} = 0.\label{zero}
\end{equation}
Again consider the region near $B$  ( $B$ is a point indicated in figure (2)
of Ref. \cite{motl2}). Since the stokes lines are equispaced near origin
(r=0), there is a difference in angle of $3\pi/2$ in complex $r$ plane, when
we pass from region $A$  to region $B$ along the stokes line. In the complex
$x$ plane, which amounts to a $3\pi$ rotation. So Bessel function will
experience a change of phase $e^{3\pi i}$ while passing from $A$ to
$B$. Taking into account the change in the Bessel function as,
\begin{equation}
 \sqrt{2\pi e^{3\pi i} \omega x}\ J_{\pm\frac{j}{2}} \left( e^{3\pi i} \omega
x \right) = e^{\frac{3\pi i}2 (1 \pm j)}\sqrt{2\pi \omega x}\
J_{\pm\frac{j}{2}} \left( \omega x \right),
\end{equation}
we can write the solution for $\Phi(x)$ to be of the form,
\begin{equation}
 \Phi (x) = \left( A_+ e^{7i\alpha_+} + B_+ e^{7i\alpha_-}\right) e^{i \omega
x} + \left( A_+ e^{5i\alpha_+} + B_+ e^{5i\alpha_-}\right) e^{-i \omega x}
\label{solution1}
\end{equation}
We now close the two asymptotic branches of the Stokes lines by a contour
along $r \sim \infty$ on which ${\rm Re}~(x) > 0$. Since we are considering
modes with ${\rm Im}~(\omega) \rightarrow \infty$, on this part of the contour
$e^{i\omega x}$ is exponentially small.  So we rely only on the coefficient of
$e^{-i\omega x}$ of Eq.  (\ref{solution1}). As the contour is completed, this
coefficient picks up a multiplicative factor given by
\begin{equation}
\frac{A_+ e^{5i\alpha_+} + B_+ e^{5i\alpha_-}} {A_+ e^{i\alpha_+} + B_+
e^{i\alpha_-}}.
\label{fac}
\end{equation}
The monodromy of $e^{-i\omega x}$ along this clockwise contour is
$e^{-\frac{\pi\omega}{k_h}}$ \cite{motl2},  where $k_h = \frac{1}{2}
f^{\prime}(r_h)$ is the surface gravity at the horizon $r_h$. Thus the
complete monodromy of the solution to the wave equation along this clockwise
contour is
\begin{equation}
\mathcal{M}(r_h)= \frac{A_+ e^{5i\alpha_+} + B_+ e^{5i\alpha_-}}{A_+
e^{i\alpha_+} + B_+ e^{i\alpha_-}} e^{-\frac{\pi\omega}{k_h}}.
\label{mono}
\end{equation}
The contour discussed above can now be smoothly deformed to a small circle
going clockwise around the horizon at $r = r_h$. Near $r = r_h$ the potential
in the wave equation approximately vanishes. From the boundary condition
Eq. (\ref{bcondition}), we see that the solution of the wave equation
Eq. (\ref{dr}) near the black hole event horizon is of the form
\begin{equation}
\Phi(x) \sim  e^{i \omega x}\,. \label{cond}
\end{equation}
The monodromy of $\Phi$ going around the small clockwise circle around the
event horizon is thus given by \cite{motl2}
\begin{equation}
\mathcal{M}(r_h) = e^{\frac{\pi\omega}{k_h}}\,.\label{mo}
\end{equation}
Since the monodromy around the two contour have to be same, we get from
Eq. (\ref{mono}) and  Eq. (\ref{mo}),
\begin{equation}
\frac{A_+ e^{5i\alpha_+} + B_+ e^{5i\alpha_-}}{A_+ e^{i\alpha_+} + B_+
e^{i\alpha_-}} e^{-\frac{\pi\omega}{k_h}} =
e^{\frac{\pi\omega}{k_h}}.\label{final}
\end{equation}
Eliminating the constants $A_+$ and $B_+$ from Eq. (\ref{zero}) and
Eq. (\ref{final}), we get
\begin{equation}
 {\rm e}^{\frac{2 \pi\omega}{k_h}} = -  \lim {j\to 0} \frac{{\rm sin}(\frac{3
\pi}{2})j}{{\rm sin}(\frac{\pi}{2})j}. \label{asyquasi}
\end{equation}
So from Eq. (\ref{asyquasi}), we can immediately write down the analytic
expression for asymptotic quasi-normal  frequency of the form,
\begin{equation}
\omega = T_h(\theta){\rm log} 3 + 2 \pi i T_h(\theta) \left ( n + \frac{1}{2}
\right ), \label{freq}
\end{equation}
where $T_h(\theta) = \frac{k_h}{2 \pi} $ is the Hawking temperature of the
non-commutative geometry inspired Schwarzschild black hole in $\hbar = 1$ unit.
%%%%%%%%%%%%%%%%%%%%%%%%%%%%%%%%%%%%%%%%%%%%%%%%%%%%%
\section{\small{\bf{Asymptotic quasi-normal mode analysis for the two horizon black hole}}}\label{two1}
%%%%%%%%%%%%%%%%%%%%%%%%%%%%%%%%%%%%%%%%%%%%%%%%%%%%%%%
We now calculate asymptotic quasi-normal modes for the case when the
non-commutative geometry inspired Schwarzschild black hole has two horizons
instead of one as in the above single horizon  analysis.  The basic
calculations are exactly similar to the one in previous section.  The only
difference comes due to the presence of  extra horizon in the  scenario. Now
the question is, how does this extra horizon modify the results of the
previous section? To get an answer to this question we need to look at
Eq. (\ref{mono}) and Eq. (\ref{mo}) where the effect of event horizon comes as
the exponential factor. So, for for the extra horizon, now we should get one
more exponential factor in both these equations. For the two horizon case now
the Eq (\ref{mono}) should be replaced by
\begin{eqnarray}
\mathcal{M}(r_h,r_{in})= \frac{A_+ e^{5i\alpha_+} + B_+ e^{5i\alpha_-}}{A_+
e^{i\alpha_+} + B_+ e^{i\alpha_-}}
e^{-\frac{\pi\omega}{k_{in}}-\frac{\pi\omega}{k_h}}\,,
\label{mono1}
\end{eqnarray}
and  Eq. (\ref{mo}) should be replaced by
\begin{equation}
\mathcal{M}(r_h,r_{in}) =
e^{\frac{\pi\omega}{k_{in}}+\frac{\pi\omega}{k_h}}.\label{1mo}
\end{equation}
This is very standard technique and can be found in the review written by 
J. Nat\'{a}rio and R. Schiappa  \cite{cardoso2}.
Now the monodromy matching between  Eq. (\ref{mono1}) and  Eq. (\ref{1mo}),
suggests that
\begin{equation}
\frac{A_+ e^{5i\alpha_+} + B_+ e^{5i\alpha_-}}{A_+ e^{i\alpha_+} + B_+
e^{i\alpha_-}} e^{-\frac{\pi\omega}{k_{in}}-\frac{\pi\omega}{k_h}} =
e^{\frac{\pi\omega}{k_{in}}+\frac{\pi\omega}{k_h}}.\label{1final}
\end{equation}
Eliminating the constants $A_+$ and $B_+$ from Eq. (\ref{zero}) and
Eq. (\ref{1final}), we get
\begin{equation}
 {\rm e}^{\frac{2 \pi\omega}{k_{in}}+\frac{2 \pi\omega}{k_h}} = -  \lim {j\to
0} \frac{{\rm sin}(\frac{3 \pi}{2})j}{{\rm
sin}(\frac{\pi}{2})j}. \label{1asyquasi}
\end{equation}
So from Eq. (\ref{1asyquasi}), we can immediately write down the analytic
expression for asymptotic quasi-normal  frequency of the form,
\begin{equation}
\omega = (T_h(\theta)+ T_{in}(\theta)){\rm log} 3 + 2 \pi i
(T_h(\theta)+T_{in}(\theta)) \left ( n + \frac{1}{2} \right ),
\label{freq1}
\end{equation}
where $T_h(\theta) = \frac{k_h}{2 \pi} $ and $T_{in}(\theta) = \frac{k_{in}}{2
\pi} $ are the Hawking temperatures of the non-commutative geometry inspired
Schwarzschild black hole at $r_{in}$ and $r_h$ respectively. In the $\theta\to
0$ limit Eq.  (\ref{1asyquasi}) reduce to the Schwarzschild value , because,
in this limit $k_{in}(\theta\to 0)= T_{in}(\theta\to 0)=\infty$. So we get
back the Schwarzschild asymptotic quasinormal modes.

So far we have considered only scalar field perturbation in our calculation.
But what happens to the Gravitational and Electromagnetic perturbations?
As we know (see appendix of review written by J. Nat\'{a}rio and R. Schiappa
\cite{cardoso2}) Gravitational perturbation can be decoupled into tensor,
vector and scalar type perturbations and Electromagnetic perturbation can be
decoupled into vector and scalar type perturbations. So we only need to
consider tensor, vector and scalar type perturbations. 
Since in our case  we assume the form of $f(r)\sim\frac{1}{r}$ near origin, the
value of $j$ in the potential $V(r)$ Eq. (\ref{rege}) for different cases are 
as follows: for scalar $j=0$, for tensor $j=0$ and for vector $j=2$.
Taking these values into consideration, we get the same result for the
asymptotic quasi-normal frequency for  all the three cases.

For the extremal black hole $M= M_0$, we can calculate the asymptotic 
quasinomal modes exactly similar to the single horizon case done in  section
3. But now the surface gravity $k_h$ and Hawking temperature $T_h$ 
will be replaced by the surface gravity and Hawking temperature
at the extremal horizon respectively.

%%%%%%%%%%%%%%%%%%%%%%%%%%%%%%%%%%%%%%%%%%%%%%%%%%%%%%%
\section{Discussion}
%%%%%%%%%%%%%%%%%%%%%%%%%%%%%%%%%%%%%%%%%%%%%%%%%%%%%%%
We have  calculated analytically the asymptotic quasi-normal frequencies for
the non-commutative geometry inspired Schwarzschild black hole in (3+1)
dimensions. Firstly, we have taken the mass of the black hole $M$ to be much
much larger than the critical mass $M_0$ \cite{nicol}. It enable us to
consider the black hole as a single horizon Schwarzschild black hole with
correction due to non-commutativity. In this situation, the real part of the
asymptotic quasi-normal frequency becomes of the form
$\frac{Re(\omega)}{T_h}=\ln(3)$, which is what we get for Schwarzschild black
hole. The effect of non-commutativity comes through the Hawking temperature
$T_h(\theta)$ and it reduces to the Schwarzschild value when the
non-commutativity goes to zero $(\theta = 0)$. Then we consider the two horizon
case and calculate its asymptotic quasi-normal frequency. We found the real
part to be again proportional to $\ln(3)$. 

Important point to note is that in our calculation we  assumed 
$r^2/4\theta$ 
very large near the origin ($r\to 0$) by making
$\theta$ very small.
This constraint can't be relaxed.  The reason is the
following:
In order to remove this constraint, we  need to consider the case 
when $\theta$ is large. That means
$r^2/4\theta$ is small near the origin ($r\to0$). For this case the upper
incomplete Gamma function should be  proportional to $r^{3/2}$, which
means $f(r)= 1- k(\theta)r^2$. $k(\theta)$ is dependent on non-commutative
parameter $\theta$. So near origin neglecting the second term in $f(r)$ for
the evaluation of tortoise coordinate $x$, we get $x\sim r$. This means we have
now two stokes lines, one along the negative imaginary axis and other
along the
positive imaginary axis on complex $r$ plane.  So now for this case if we want
to calculate the asymptotic quasi-normal mode we need to calculate the phase
factor of the wave function going from region A to region B (figure 2
of Ref. \cite{motl2}). This phase factor is $\pi$ (both $r$ and $x$ have
the same phase factor in this case). Taking into account  this phase
factor if
we follow the calculation of   section 3, we
will get a multiplicative factor $1$ instead of Eq. (\ref{fac}). 
This will lead
to $1$ in the r.h.s of Eq. (\ref{asyquasi}), which means  
asymptotic quasi-normal
frequency is  proportional to $0$!. But this
contradicts
our assumption that the imaginary part of frequency is very large.
We can't resolve it using monodromy technique. Perhaps more careful analysis
should be made in order to relax the constraint made on $r^2/4\theta$.
Finally generalization of the idea
\cite{nicol} to any dimension $(d>3)$ and study of its quasi-normal modes would
be interesting. Study of low damping quasi-normal modes of this kind of black
hole is also an open problem. We hope to perform these problems in future.

%%%%%%%%%%%%%%%%%%%%%%%%%%%
\subsubsection*{Acknowledgments}
%%%%%%%%%%%%%%%%%%%%%%%%%%%
We thank  Vitor Cardoso and Ricardo Schiappa for comments on manuscript and
valuable suggestions.
%discussions.
%%%%%%%%%%%%%%%%%%%%%%%%%%%


\begin{thebibliography}{[W]}
%%%%%%%%%%%%%%%%%%%%%%%%%%%
\bibitem{press1} W. H. Press, Astrophys. J {\bf L105} 170 (1971).

\bibitem{press2} H. P. Nollert, Class. Quant. Grav. {\bf 16} R159 (1999). 


\bibitem{press3} K. D.  Kokkotas and B.G. schmidt, Living Rev. 
Rel. {\bf 2} 2 (1999).


\bibitem{hod} S. Hod, Phys. Rev. Lett. {\bf 81}  4293 (1998).

\bibitem{dreyer} O. Dreyer, Phys. Rev. Lett. {\bf 90} 081301 (2003).

\bibitem{motl1} L. Motl, Adv. Theor. Math. Phys. {\bf 6} 1135 (2003).

\bibitem{motl2} L. Motl and A. Neitzke, Adv. Theor. Math. Phys. {\bf 7}  307
(2003).

\bibitem{cardoso1} V. Cardoso, J. Nat\'{a}rio and R. Schiappa, J.
Math. Phys. {\bf 45} 4698 (2004).

\bibitem{cardoso2} J. Nat\'{a}rio and R. Schiappa, Adv. Theor. Math. Phys. 
{\bf 8} 1001 (2004).

\bibitem{cardoso3} S. K. Chakrabarti, K. S. Gupta, Int. J. Mod. Phys. {\bf A21}
 3565 (2006). 

\bibitem{cardoso4} S. Das and S.
Shankaranarayanan, Class. Quant. Grav. {\bf 22}, L7 (2005).


\bibitem{ko1} R. A. Konoplya,  Phys. Rev. {\bf D73} 024009 (2006).

\bibitem{ko2} R. A. Konoplya and  A.V. Zhidenko, Phys. Lett. {\bf B609} 
377 (2005).
\bibitem{ko3} R. A. Konoplya and  A. Zhidenko, JHEP 0406 037 (2004).

\bibitem{witten1} E. Witten, Nucl. Phys. {\bf B460} 33 (1996). 

\bibitem{witten2} M. R. Douglas
and C. M. Hull, J. High Energy Phys. 02 008 (1998).

\bibitem{witten3} C. Chu and P.  Ho,
Nucl. Phys. {\bf B550} 151 (1999). 

\bibitem{witten4} C. Chu and P.  Ho,
Nucl. Phys. {\bf B568} 447 (2000). 

\bibitem{witten5} N.  Seiberg and
E. Witten, J. High Energy Phys. {\bf 09} 032  (1999).

\bibitem{anto} I. Antoniadis, N. Arkani-Hamed, S. Dimopoulos and G.  R. Dvali,
Phys. Lett. {\bf B436} 257 (1998).

\bibitem{gru1} A. Gruppuso, J. Phys. {\bf A38} 2039 (2005). 

\bibitem{gru2} P. Nicolini,
J. Phys. {\bf A38} L631 (2005).

\bibitem{nasseri1} F. Nasseri, Gen.Rel.Grav. {\bf 37} 2223 (2005).
 
\bibitem{nasseri2} F. Nasseri, Int. J. Mod. Phys. {\bf D15} 1113 (2006).
 
\bibitem{nasseri3} Xin-zhou Li, hep-th/0508128.

\bibitem{nicol}P. Nicolini, A. Smailagic and E. Spallucci, Phys.Lett. {\bf
B632} 547 (2006).

\bibitem{nicoli}P. Nicolini,  hep-th/0510203.

\bibitem{regge}  T. Regge and J. A. Wheeler, Phys. Rev. {\bf 108} 1063 (1957).


%%%%%%%%%%%%%%%%%%%%%%%%%%%%%%%%%%%%%%%%%%%%%%%%%%%%%%%%%%

%\bibitem{ads} G. T. Horowitz, V. E. Hebuny, Phys. Rev. {\bf D62},
%024027 (2000).

%\bibitem{cardoso} V. Cardoso, J. Nat\'{a}rio and R. Schiappa, J.
%Math. Phys. {\bf 45}, 4698 (2004).


%\bibitem{natario} J. Nat\'{a}rio and R. Schiappa, hep-th/0411267.



\end{thebibliography}
\end{document}